\documentclass[conference]{IEEEtran}
\IEEEoverridecommandlockouts

\renewcommand{\baselinestretch}{.9} 

\usepackage{cite}
\usepackage{amsmath,amssymb,amsfonts}
\usepackage{algorithmicx}
\usepackage{algorithm}
\usepackage[noend]{algpseudocode}
\usepackage{graphicx}
\usepackage{textcomp}
\usepackage{caption}
\usepackage{subcaption}

\usepackage{fancyhdr}
\author{

\IEEEauthorblockN{Amir Ali-Pour}
\IEEEauthorblockA{
\textit{ÉTS Montréal / Université du Québec}\\
amir.ali-pour@etsmtl.ca}
\\
\IEEEauthorblockN{laya Samizadeh}
\IEEEauthorblockA{
\textit{ÉTS Montréal / Université du Québec}\\
laya.samizadeh.1@ens.etsmtl.ca}
\and
\IEEEauthorblockN{Sadra Bekrani}
\IEEEauthorblockA{
\textit{Islamic Azad University of Bojnord}\\
sadra.bekrani@iau.ir}
\\
\IEEEauthorblockN{Julien Gascon-Samson}
\IEEEauthorblockA{
\textit{ÉTS Montréal / Université du Québec}\\
julien.gascon-samson@etsmtl.ca}
\thanks{}}

\usepackage{xcolor}
\def\BibTeX{{\rm B\kern-.05em{\sc i\kern-.025em b}\kern-.08em
    T\kern-.1667em\lower.7ex\hbox{E}\kern-.125emX}}
\begin{document}

\title{Towards a Distributed Federated Learning Aggregation Placement using Particle Swarm Intelligence}


\maketitle

\begin{abstract}
Federated learning has become a promising distributed learning concept with extra insurance on data privacy. Extensive studies on various models of Federated learning have been done since the coinage of its term. One of the important derivatives of federated learning is hierarchical semi-decentralized federated learning, which distributes the load of the aggregation task over multiple nodes and parallelizes the aggregation workload at the breadth of each level of the hierarchy. Various methods have also been proposed to perform inter-cluster and intra-cluster aggregation optimally. Most of the solutions nonetheless require monitoring the nodes' performance and resource consumption at each round, which necessitates frequently exchanging systematic data. To optimally perform distributed aggregation in SDFL with minimal reliance on systematic data, we propose Flag-Swap, a Particle Swarm Optimization (PSO) method that optimizes the aggregation placement according only to the processing delay. Our simulation results show that PSO-based placement can find the optimal placement relatively fast, even in scenarios with many clients as candidates for aggregation. Our real-world docker-based implementation of Flag-Swap over the recently emerged FL framework shows superior performance compared to black-box-based deterministic placement strategies, with about $43\%$ minutes faster than random placement, and $32\%$ minutes faster than uniform placement, in terms of total processing time.

\end{abstract}

\begin{IEEEkeywords}
Distributed Systems, Federated Learning, Aggregation, Task Placement, Swarm Intelligence, Black-box Optimization
\end{IEEEkeywords}

\section{Introduction}

Federated Learning (FL) has emerged as a revolutionary approach to distributed machine learning within Internet of Things (IoT) ecosystems \cite{nguyen2021federated, zhang2022federated}. With the rapid expansion of IoT devices, vast amounts of decentralized data are being generated at the network edge, posing significant challenges for traditional centralized learning methods. These conventional approaches require transferring data to central servers, leading to high bandwidth costs, increased latency, and serious privacy concerns. In contrast, FL facilitates collaborative model training directly on edge devices, allowing them to contribute to a shared global model without transmitting raw data \cite{lim2020federated}. This capability is particularly beneficial for IoT environments, where efficient bandwidth utilization, enhanced privacy, and real-time responsiveness are crucial. By minimizing data transmission, preserving data privacy, and optimizing edge computational resources, FL effectively overcomes key limitations of centralized learning \cite{ji2023joint, guo2021efficient}.

The key part of FL ecosystems is the aggregators, which are nodes that accumulate model parameters or their gradients from the individual nodes and accumulate them using various aggregation methods. The aggregation yields a new set of model parameter values that speculatively represent the learned features from all the contributing nodes' data. Various FL schematics exist, which depend heavily on the underlying network topology. There are three main categories: \textbf{1) Central FL (CFL)} is the conventional FL model which is based on the client/server communication model, and follows a star topology, wherein one central unit (i.e., parameter server or aggregation server) is responsible for performing the global model update, and thus all the contributing clients would send their model parameters to that central unit. \textbf{2) Fully Decentralized FL (DFL)} is a model that follows a P2P communication method, and no central unit is dedicated to aggregation. Instead, model parameters are aggregated after each hop at the destination client machine. \textbf{3) Semi-Decentralized FL (SDFL)} is a hybrid model between the CFL and DFL, wherein the aggregation load is spread down onto multiple machines, and the aggregator machines either synchronously or asynchronously deliver the aggregation with mutual agreement on the global model updates. This FL model promises parallelism while avoiding a single point of failure, given that the aggregation is distributed across multiple nodes, thus the system is more resilient to node failures or connectivity issues. One of the known SDFL models is the Hierarchical SDFL, in which the aggregation is spread not only specially at the breadth of each hierarchy level but also temporally between each hierarchy level \cite{liu2020client}. On top of the advantages of SDFL, Hierarchical SDFL promises scalability, reduced computation bottleneck, and better adaptation to system constraints. From hereon we refer to Hierarchical SDFL as SDFL.

One of the key challenges in SDFL is to find a set of suitable machines as aggregators. The criteria behind choosing an aggregator machine can be bound to several parameters, including key systematic parameters such as the availability of the machine, its computation resources, and communication bandwidth. Several methods have been proposed \cite{luo2020hfel} that use different sets of parameters to create the criterion for developing or optimizing the search for a suitable aggregation site. Nonetheless, most of these methods require the contributing clients to inform the coordinator of their internal performance, which could impose challenges such as network congestion if such data is requested frequently, or violate the privacy of the contributing clients. In contrast to such methods, task placement methods that follow black-box system optimization exist which have seldom been practiced for SDFL. Given that, one can incorporate such an optimization, and in turn, guarantee an optimal placement of aggregation while avoiding transmission and additional processing of the client machines' internal performance for a supervised optimization. In this paper, we set the goal to investigate the efficacy of using such optimizers. Specifically, we propose using the particle swarm optimization (PSO) method to progressively improve the placement of aggregation. We show that we can improve the placement of aggregation with PSO with regard only to the global processing delay at each FL round. We also demonstrate that PSO imposes marginal computational complexity, given if a suitable FL framework is used that supports Hierarchical FL implementation, making the optimizer a suitable candidate for constrained systems at the edge. Following is the list of contributions we deliver:
\begin{itemize}
    \item A black-box PSO-based aggregation placement for SDFL 
    \item Evaluation of the efficacy of the optimizer in various simulated SDFL scenarios with different numbers of clients and varied depth and width in the hierarchy model.
    \item Evaluation of the efficacy of the optimizer in a real SDFL ecosystem based on MQTT communication deployed on docker containers.
    \item Comparison with random placement and uniform placement based on round-robin algorithm.
\end{itemize}

The rest of the paper is as follows: Section II presents the motivation behind employing a black-box optimizer in an SDFL based on the Publish/Subscribe communication model. Section III explains the key features and the mechanism of the proposed optimizer for aggregation placement. Section IV describes the experimental setup and the experimental results using both simulation and real-world deployment. Section V is a discussion of the related works. Section VI concludes the paper.

\section{Motivation}
A commonplace communication model for SDFL would be the Client/Server model similar to CFL \cite{beltran2023decentralized}. While this model is effective for systems with substantial computational resources and stable network connections, it is not well-suited for environments with resource-constrained devices, such as those found in IoT networks. In such scenarios, dynamic role management, where devices alternate as aggregators to mitigate overload and device exhaustion, becomes essential. Implementing this in a client/server architecture would require complex mechanisms for dynamic role assignment. Alternatively, a fully decentralized peer-to-peer (P2P) approach can ensure that aggregation roles are distributed effectively, though it incurs a training time overhead due to sequential communication. 


Lately, a proposition was made to use the Publish/Subscribe communication model instead of Client/Server \cite{Ali-Pour2025SDFLMQ}. They integrate such a service, which only requires a broker at the edge to disseminate the model updates, while the FL-specific roles are delegated to the devices that need the ML services. Therefore, at the edge, role association would be as general as just a message disseminator which does not need any adaptation to the FL process. For instance, if an MQTT broker is running as a service on an edge server, we can connect to that and establish the FL roles among the devices connected to the broker. This would in turn help set up the framework faster and with reduced cost of installment.

\begin{figure}[t]
    \centering
    \includegraphics[width=.9\linewidth]{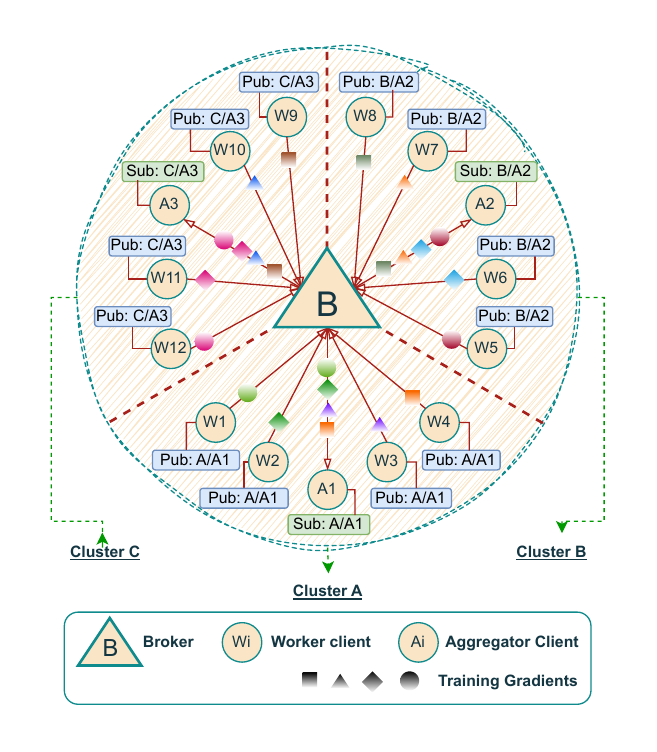}
    \caption{Overview of Parameter sharing for aggregation using Pub/Sub communication in a Clustered Semi-Decentralized Federated Learning Topology.}
    \label{fig:overview}
\end{figure}


SDFL over MQTT is a promising practice, that provides simplified orchestration, avoids single point of failure, and increases redundancy. Role association and role management in SDFL over MQTT as described in \cite{Ali-Pour2025SDFLMQ} can be managed relatively easily compared to SDFL implementations using other FL frameworks. This is because in SDFLMQ, FL roles are associated to topics. Following that, candidates for each role can choose to subscribe to their role's topic, and clients that want to communicate to a node with a specific role, can publish to that role's topic. The simplicity of role management in this SDFL model helps save time and energy in changing FL's actor roles during the FL process. Additionally, it opens more room to develop more sophisticated optimization algorithms.

Regarding load balancing and task scheduling, numerous techniques can be used to solve this problem. However, in the context of SDFLMQ as described in \cite{Ali-Pour2025SDFLMQ}, one can notice that there is anonymity in the contribution of clients to the FL process. Meaning that clients do not share any information about their internal status to register their candidacy for aggregation with the coordinator. This anonymity in turn enables further expandability and upholds clients' data privacy. Nonetheless, as mentioned earlier, most of the load-balancing techniques for SDFL need to process clients' systematic data to choose suitable sites for aggregation. To be able to perform aggregation placement without requiring such data, one can think of incorporating black-box-based optimization techniques. These techniques can perform optimization with only some macro measurements of the entire system such as the total processing delay, or total energy consumption. Solutions that fall into black-box optimization could be evolution strategies, Bayesian optimization, ant colony optimization, genetic algorithm (GA), swarm intelligence, reinforcement learning, etc \cite{meunier2021black, boveiri2020performance,auger2009experimental}.


While most of these algorithms are potentially applicable to solving the aggregation placement in SDFL, PSO can be found the most potential, mainly due to its convergence speed. Several studies compared PSO to other algorithms such as GA, and concluded that PSO in turn has better performance and convergence whereas GA yields premature convergence \cite{boveiri2020performance}. Given that we aim to optimize the aggregation placement with regards to total processing delay, better performance in the optimizer algorithm of course can lead to better placement which in SDFL would lead to lowered total processing delay. Fast convergence also means that we would go through fewer trials until we reach a status where all suggestions (i.e., particles in PSO) lead to a local/global best placement. Given that, it is justified to implement a placement optimizer in SDFL using PSO. In the following, we explain our aggregation placement optimizer based on PSO for SDFL over MQTT.

\begin{figure}[t]
    \centering
    \includegraphics[width=.9\linewidth]{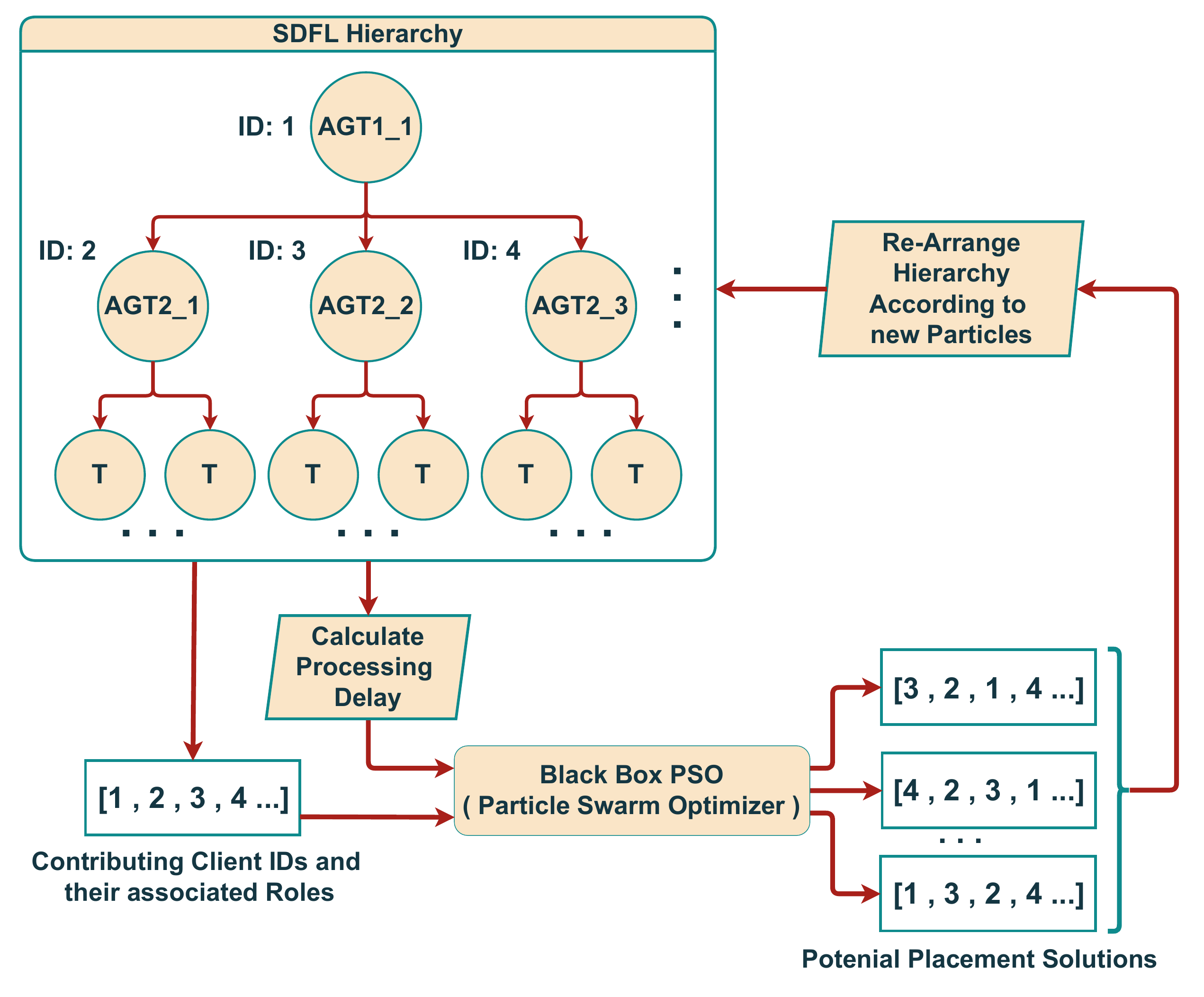}
    \caption{Proposed PSO-based aggregation placement in SDFL.}
    \label{fig:pso}
\end{figure}

\section{Proposed Method}

In our black‐box PSO approach, clients do not share their internal performance metrics. The coordinator records the processing time of each round and computes the processing delay by subtracting the round's start time from the round's ending time. This in turn elevates the necessity of each client informing the coordinator of the internal performance or processing delay, thus significantly reducing the communication load while preserving the privacy of each client. The core objective of our method is to progressively minimize the total processing delay (TPD) of the FL rounds through the PSO optimization loop. Fig. \ref{fig:pso} shows the general overview of agg placement in SDFL using PSO. 

To achieve optimal placement, we update the clients' roles by efficiently arranging them as either trainers or aggregators before the beginning of each round. By leveraging the global search capabilities of PSO, the method explores a vast solution space of possible client arrangements and identifies configurations that lead to reduced latency, critical for scalability and real-time performance. Thus, at each round, after computing the processing delay of the previous round, PSO suggests a new arrangement according to its particles. The PSO particles are also updated after each PSO fitness round according to the local and global particle fitness values.

\subsection{Particle Swarm Optimization for Client Placement}
We employ PSO to optimize the assignment of clients to aggregator roles within the hierarchy. In this formulation:
\begin{itemize}
    \item \textbf{Particle Representation:} Each particle represents a potential arrangement solution. Each element in the vector is a client ID assigned to an aggregator slot.
    \item \textbf{Swarm:} A population of \( P \) particles explores the solution space.
    \item \textbf{Velocity:} Each particle has a velocity vector that dictates how its position changes in each iteration.
\end{itemize}

\subsection{Fitness Function}
The quality of a client arrangement is evaluated using a fitness function based on the Total Processing Delay (TPD). The fitness \( f \) of an arrangement is:
\begin{equation}
    f = -T
\end{equation}
where $T$ is the TPD of the corresponding FL round. By maximizing \( f \), we effectively minimize \( T \). This formulation captures the bottleneck effect at each hierarchy level, ensuring that the arrangement balances the computational load across the hierarchy.

\subsection{Optimization Loop}

The optimization loop in PSO for aggregation placement in SDFL is the following: 
\begin{itemize}
    \item A swarm of \( N \) particles is initialized (e.g., \( N = 10 \)).
    \item The initial position of each particle is a random permutation of client IDs assigned to aggregator roles. 
    \item Initial velocities are set to zero.
    \item The personal best position of each particle is its initial position, and the global best position is the position yielding the highest initial fitness.
\end{itemize}

The optimization loop steps are the following:
\begin{enumerate}
    \item \textbf{Velocity Update:}
    \begin{equation}
        v_{i}^{t+1} = w \cdot v_{i}^t + c_1 \cdot r_1 \cdot (p_{i} - x_{i}^t) + c_2 \cdot r_2 \cdot (g - x_{i}^t)
    \end{equation}
    where:
    \begin{itemize}
        \item \( v_{i}^t \): Velocity vector of particle \( i \) at iteration \( t \).
        \item \( x_{i}^t \): Position of particle \( i \)at iteration \( t \).
        \item \( p_{i} \): Personal best position of particle \( i \).
        \item \( g \): Global best position.
        \item \( w \): Inertia weight (e.g., 0.01).
        \item \( c_1 \): Cognitive coefficient (e.g., 0.01).
        \item \( c_2 \): Social coefficient (e.g., 1).
        \item \( r_1, r_2 \): Random numbers in [0, 1].
    \end{itemize}
    Velocity components are clamped to the interval \([-V_{\max}, V_{\max}]\), where :
    \begin{equation}
        V_{\max} = \max\left(1, \, D \times \textit{velocity\_factor}\right)
    \end{equation}
    and \( D \) is the number of dimensions in the search space. For example, a typical value is \(\text{velocity\_factor} = 0.1\).
    \item \textbf{Position Update:}
    The new position is computed as:
    \begin{equation}
        x_{i}^{t+1} = (x_{i}^t + v_{i}^{t+1}) \mathbin{\%} \textit{client\_count}
    \end{equation}
    Duplicates are resolved by incrementing until a unique client ID is found.
    \item{\normalfont\textbf{Hierarchy Rearrangement:}}
    After updating a particle's position:
    \begin{itemize}
        \item Clients are reassigned aggregator roles based on the updated particles.
        \item Remaining clients are assigned trainer roles from a buffer of available labels.
    \end{itemize}

    \item{\normalfont\textbf{Iteration and Convergence:}}
    The algorithm iterates for \( M \) steps, updating personal and global bests when better fitness values are found. This usually happens when the TPD value is converged to a minimum value. The final global best position represents the optimal client placement. Algorithm \ref{pso_algo} shows the iterative process of swarm optimization.
\end{enumerate}

\begin{algorithm}[t]
\caption{PSO Algorithm for SDFL}
\label{pso_algo}
\begin{algorithmic}[]
\State \textbf{Inputs:}
\State \quad $DEPTH$, $WIDTH$, $pop_n$, $max\_iter$, $iw$, $c1$, $c2$, $velocity\_factor$

\State \textbf{Initialization:}
\State \quad Generate hierarchy with aggregators and trainers
\State \quad Create $pop_n$ particles with positions (client assignments)
\State \quad Compute initial fitness for each particle

\State \textbf{Main Loop:}
\For{$iteration \gets 1$ to $max\_iter$}
    \For{each particle $p$}
        \State Update velocity using $iw$, $c1$, $c2$
        \State Update position based on velocity
        \State Rebuild hierarchy with new assignments
        \State Compute new fitness
        \If{new fitness better than $pbest$}
            \State Update $pbest$
        \EndIf
        \If{new fitness better than $gbest$}
            \State Update $gbest$
        \EndIf
    \EndFor
\EndFor

\State \textbf{Processing\_Fitness Function:}
\State \quad Traverse hierarchy bottom-up
\State \quad Compute memory consumption and delays per level
\State \quad Sum maximum delays across levels
\State \quad Return fitness, total delay
\end{algorithmic}
\end{algorithm}


\begin{figure*}[t]
     \centering
     \begin{subfigure}[]{0.325\textwidth}
         \centering
         \includegraphics[width=\textwidth]{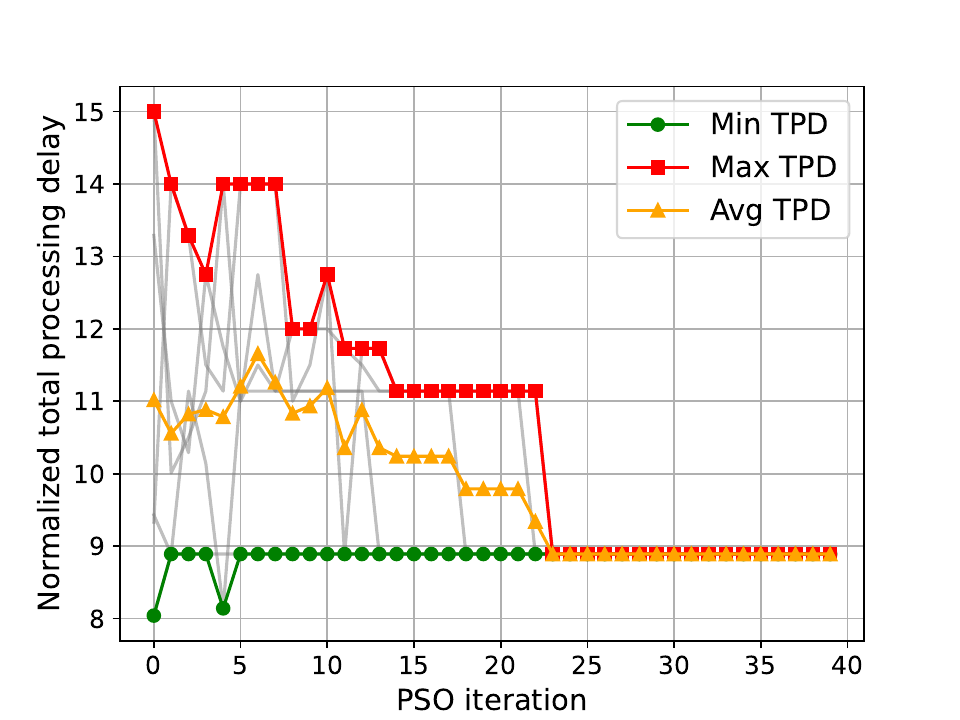}
         \caption{Client number 81, DEPTH:3, WIDTH: 5, Num particles:5}
         \label{fig:three sin x}
     \end{subfigure}
     \hfill
     \begin{subfigure}[]{0.325\textwidth}
         \centering
         \includegraphics[width=\textwidth]{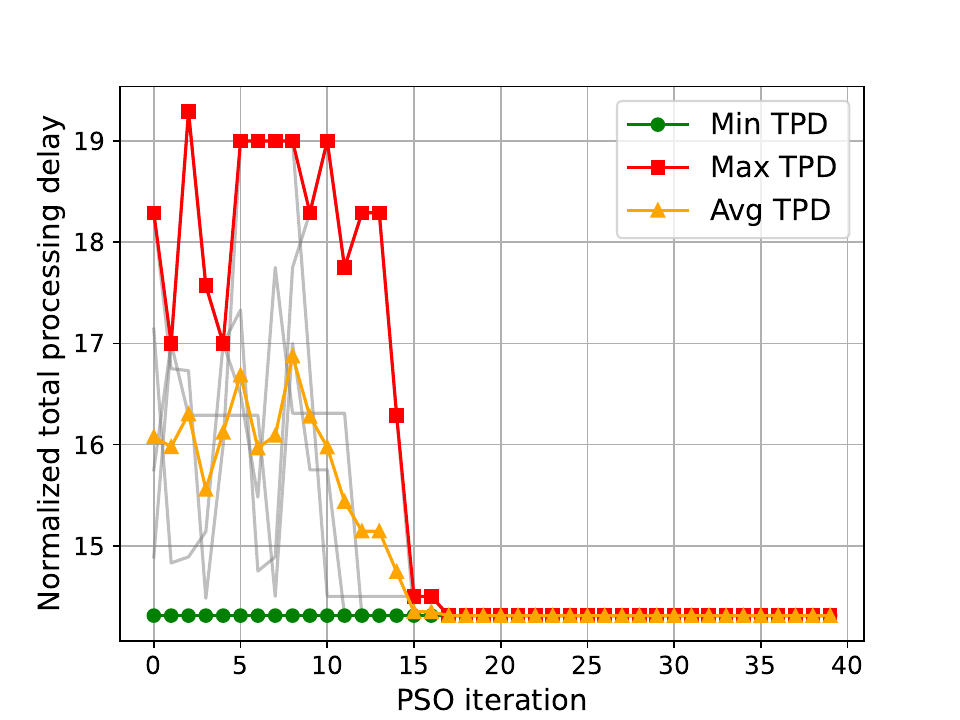}
         \caption{Client number 406, DEPTH:4, WIDTH: 5, Num particles:5}
         \label{fig:five over x}
     \end{subfigure}
     \begin{subfigure}[]{0.325\textwidth}
         \centering
         \includegraphics[width=\textwidth]{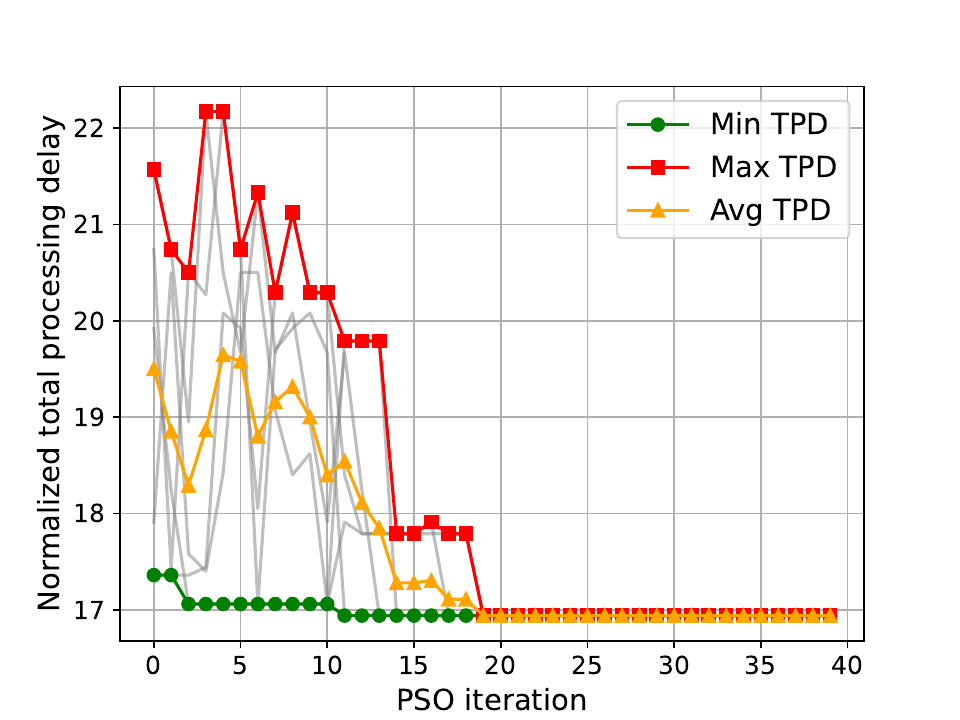}
         \caption{Client number 853, DEPTH:5, WIDTH: 4, Num particles:5}
         \label{fig:y equals x}
     \end{subfigure}
     \hfill
     \begin{subfigure}[]{0.325\textwidth}
         \centering
         \includegraphics[width=\textwidth]{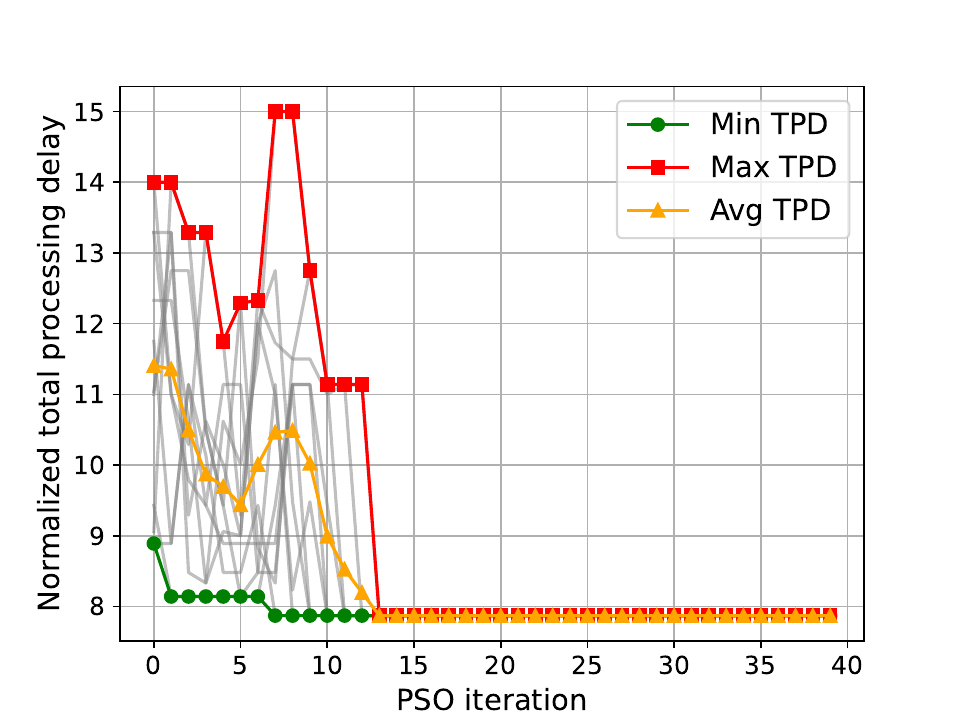}
         \caption{Client number 81, DEPTH:3, WIDTH: 5, Num particles:10}
         \label{fig:three sin x}
     \end{subfigure}
     \hfill
     \begin{subfigure}[]{0.325\textwidth}
         \centering
         \includegraphics[width=\textwidth]{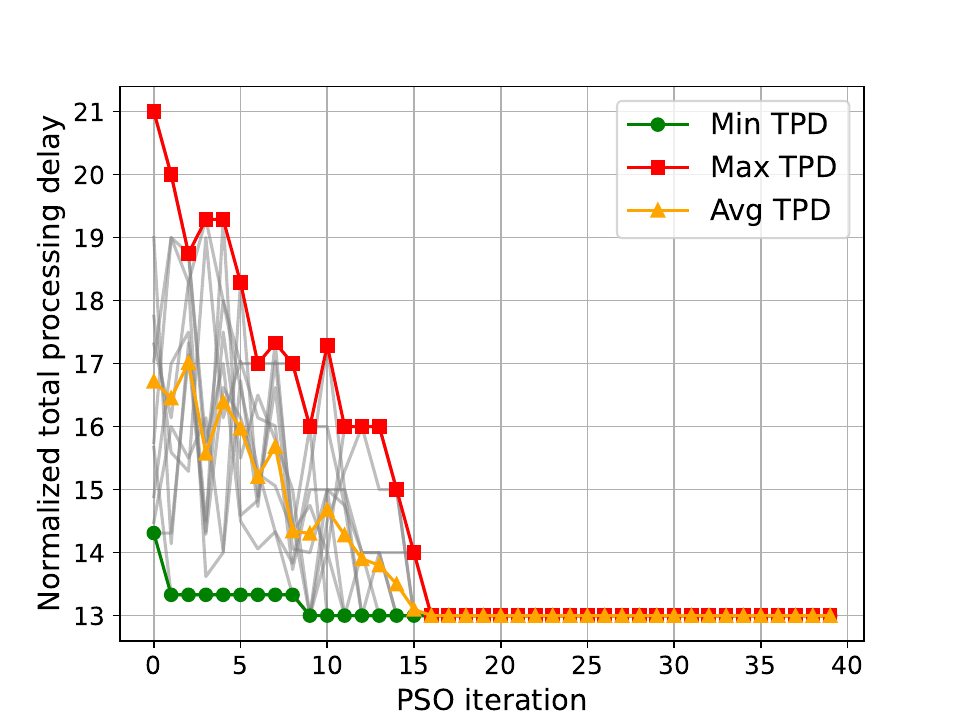}
         \caption{Client number 406, DEPTH:4, WIDTH: 5, Num particles:10}
         \label{fig:five over x}
     \end{subfigure}
     \begin{subfigure}[]{0.325\textwidth}
         \centering
         \includegraphics[width=\textwidth]{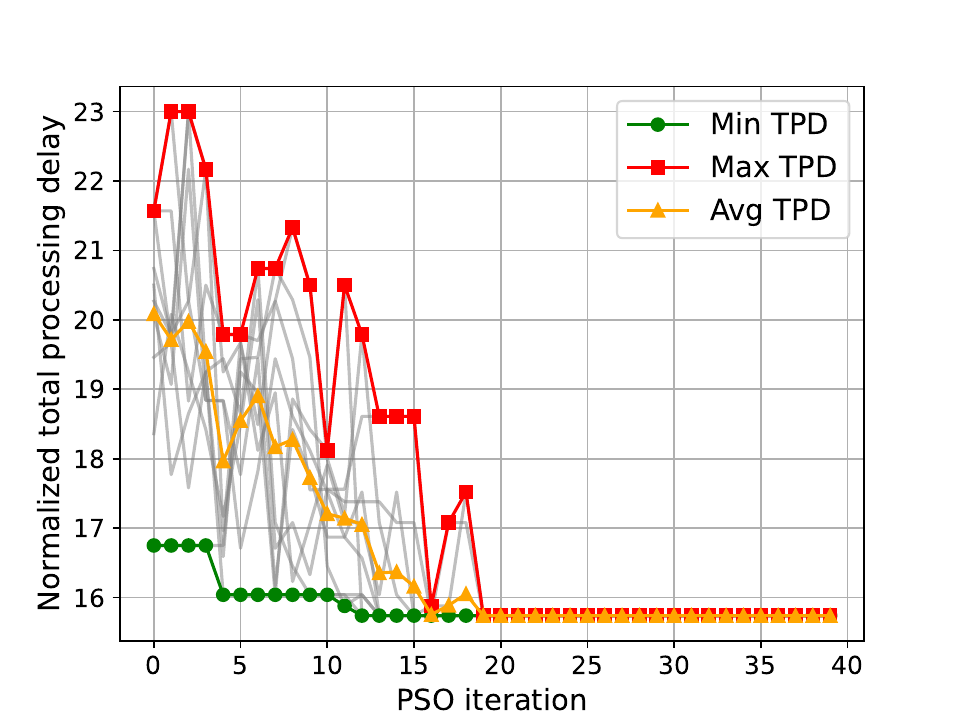}
         \caption{Client number 853, DEPTH:5, WIDTH: 4, Num particles:10}
         \label{fig:y equals x}
     \end{subfigure}
     \hfill
        \caption{Simulation results of PSO optimization in aggregation placement in SDHFL.}
        \label{fig:sim_measurements}
\end{figure*}

\section{Experimental Setup \& Results}
\subsection{Simulation Model}
We model the FL system as a hierarchical tree with a depth \( D \) and a width \( W \). The hierarchy comprises clients with two distinct roles:
\begin{itemize}
    \item \textbf{Aggregators (Agtrainers):} Nodes responsible for aggregating model updates from their child clients. Each aggregator maintains a processing buffer containing its children, which can be either trainers (for layer \(D - 1\)) or other aggregators.
    \item \textbf{Trainers:} Leaf nodes that perform local model training and send updates to their parent aggregators.
\end{itemize}
Each client \( c_i \) is defined by the following attributes:
\begin{itemize}
    \item Memory capacity \( \textbf{memcap}_i \): The memory capacity of the client.
    \item Model data size \( \textbf{mdatasize}_i \): The size of the model data processed by the client (fixed at 5 units in this study).
    \item Processing speed \( \textbf{pspeed}_i \): The computational speed of the client, randomly assigned between 5 and 15 units.
    \item Client ID \( \textbf{client\_id}_i \): A unique identifier for the client.
\end{itemize}
The hierarchy is constructed recursively starting from a root aggregator at level 0. For each level \( l \) (where \( 0 \leq l < D-1 \)), an aggregator has \( W \) child aggregators at level \( l+1 \). At the leaf level (\( l = D-1 \)), each aggregator is assigned several trainers (e.g., 2 in our simulation model). The total number of aggregator positions, or dimensions, is computed as:
\begin{equation}
\textit{dimensions} = \sum_{i=0}^{D-1} W^i
\end{equation}
This represents the number of slots in the hierarchy where clients can be assigned as aggregators. The fitness function $f$ is implemented as the following: We first use Breadth-First Traversal (BFT) to organize the hierarchy into levels, starting from root. Then, we calculate the TPD by processing these levels from the bottom (leaf nodes) to the top (root). For each level, we determine the maximum cluster delay among all aggregators, and the TPD is the sum of these maximum delays across all levels.
For an aggregator \( a \), the cluster delay \( d_a \) is defined as:

\begin{equation}
    d_a = \frac{\textit{mdatasize}_a + \sum_{c \in \textit{children}(a)} \textit{mdatasize}_c}{\textit{pspeed}_a}
\end{equation}

\noindent
where \( \text{children}(a) \) denotes the set of clients in \( a \)'s processing buffer. The total processing delay (TPD) \( T \) is:
\begin{equation}
    T = \sum_{\textit{levels}} \max_{a \in \textit{level}} d_a
\end{equation}

\noindent

\subsection{Simulation setup \& results}

A simulation was implemented featuring an SDFL system with a hierarchical structure of depth $N \in \{3,4,5\}$ and width $M \in \{4,5\}$, constructed via breadth-first traversal to ensure balanced role distribution. Clients within this hierarchy are categorized as either aggregators or trainers. Each simulated client node has a processing buffer that is used to keep their child nodes within an array, and if those child nodes are also aggregators, they maintain their non-empty processing buffers. Trainer nodes also have processing buffers, which remain empty. Trainers retain these buffers because their role might change later, potentially transitioning into an aggregator position. Each client is assigned random attributes, including memory capacity $10 < m < 50$, processing speed $5 < ps < 15$ units, and a uniform model data size fixed at $5$. The PSO-based role assignment changes the position of simulated client nodes in the hierarchy which in turn affects the TPD. Note that the Total Processing Delay (TPD) is calculated as the sum of the maximum cluster delays in each level of the hierarchical structure. The role adjustments lead to minimizing the TPD across the system.

Optimization of client role assignments is achieved through PSO utilizing a swarm of $P \in \{5,10\}$ particles, each representing a potential configuration of the hierarchical structure. Note that each particle indicates the position of the aggregator clients. Trainer clients will be assigned randomly as the terminal node to the aggregators. The PSO algorithm is configured with an inertia weight of $0.01$ to favor exploitation, a cognitive coefficient (c1) of $0.01$ for stability with the small swarm size, and a social coefficient (c2) of $1$ to emphasize the influence of the global best solution. It iterates for $100$ generations, with a velocity factor of $0.1$.


Results of the aggregation placement using PSO in simulated SDFL are shown in Fig. \ref{fig:sim_measurements}. Each plot shows the normalized TPD with respect to PSO iterations. Grey curves show the processing delay per PSO particle, and the red, green, and orange curves show the worst, best, and average processing delay at each iteration step, respectively. The key observation her is the convergence of TPD. As expected, PSO particles manage to lead the TPD to a minimum value, up to a point where all the particles suggest the same placement which results in the global minimum TPD. The convergence of all particles to one placement is needed, since at each FL round when a particle is given for a new placement, it is not assured if the particle will lead to a new minimum TPD. The only way is to test the particle and calculate the TPD after the global model is yielded for that round. Once the particles converge, we can ensure that the optimizer has searched the potential placements in the search space while heuristically progressing toward minimizing the TPD. 

Moreover, we can also see that PSO adapts well to the increasing number of clients, even though knowing that the dimensionality of the particles in cases with large numbers of clients would be high. We can see this by comparing Fig. \ref{fig:sim_measurements} (a) with Fig. \ref{fig:sim_measurements} (b) and Fig. \ref{fig:sim_measurements} (c), and Fig. \ref{fig:sim_measurements} (d) with Fig. \ref{fig:sim_measurements} (e) and Fig. \ref{fig:sim_measurements} (f). The last observation is the effect of increasing the number of particles. We can see that a larger number of particles can potentially result in finding a better placement leading to an even lower TPD value. This can be seen in comparing the results in Fig. \ref{fig:sim_measurements} (a) with Fig. \ref{fig:sim_measurements} (d), or Fig. \ref{fig:sim_measurements} (b) with Fig. \ref{fig:sim_measurements} (e), or Fig. \ref{fig:sim_measurements} (c) with Fig. \ref{fig:sim_measurements}(f).

\subsection{Docker-based setup \& results}
\begin{figure}[t]
    \centering
    \includegraphics[width=1\linewidth]{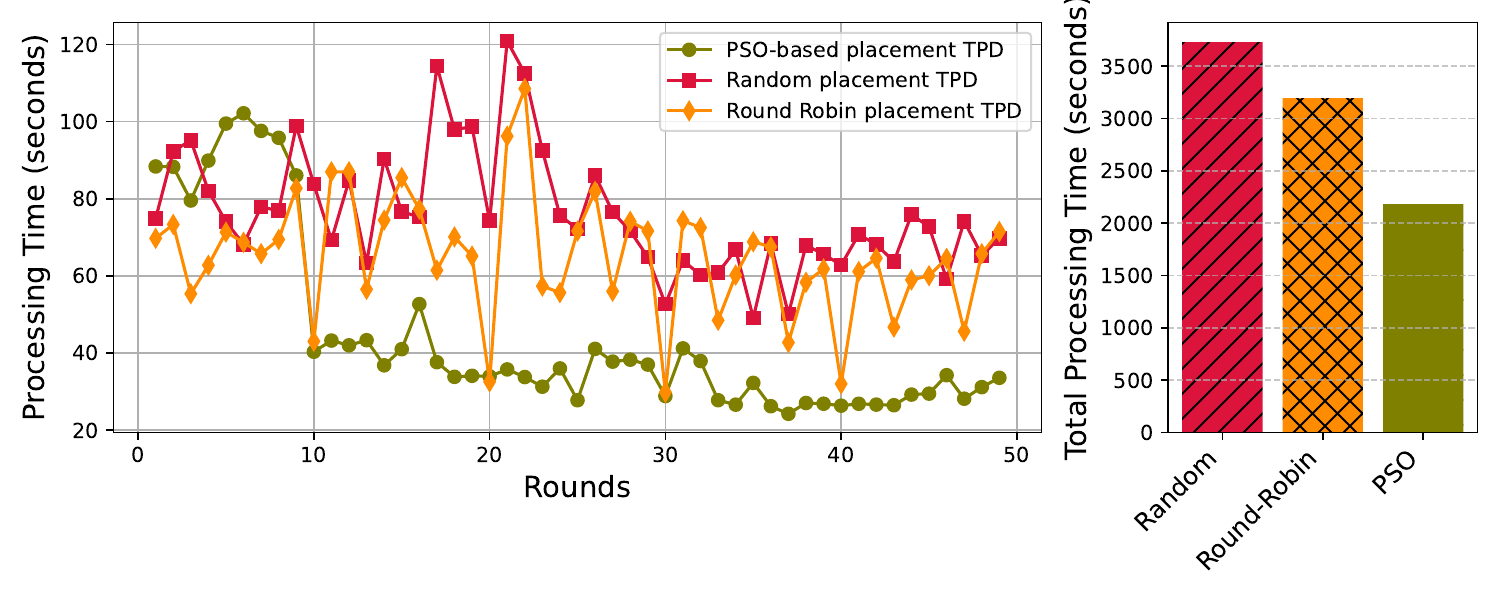}
    \caption{Comparing aggregation placement using Random, PSO-based, and Round-Robin-based placement in SDFLMQ}
    \label{fig:docker_measurements}
\end{figure}

To evaluate the applicability of PSO and it's potential use in real systems, we integrated our implementation into the SDFLMQ framework's code which is publicly available at \cite{SDFLMQsource}, and compared the performance of our method with the builtin placement strategies including random placement and uniform round-robin-based placement. We created one scenario, including 10 docker-container clients, with one client having $2Gb$ dedicated memory and $3$ dedicated cores, two clients with $1Gb$ dedicated memory, $1Gb$ capacity for memory swap, and $1$ core each, and seven clients with $64Mb$ dedicated memory, $2Gb$ capacity for memory swap, and $1$ dedicated core each. We gave a multi-layer perceptron model to each client, with $1.8$ million parameters, and about $30Mb$ of size in json format, which is the format used in SDFLMQ to write the model parameters in and transmit in-between SDFLMQ nodes. We run the scenario for $50$ rounds, and recorded the processing delay at each round, and the total processing delay after 50 rounds. Fig. \ref{fig:docker_measurements}  shows the processing delay, per round for the three placement strategies including random placement, uniform round-robin placement, and PSO-based placement. As can be seen, PSO-based placement was able to converge after the $10^{th}$ round. After the convergence, PSO-based placement shows between $20$ seconds to $30$ seconds faster processing time per round, compared to random-based and uniform-based placements. The total processing time in PSO-based placement also is significantly better, leading to around $30$ minutes faster than in random-based placement, and around $20$ minutes faster than in uniform-based placement.

Overall, the evaluation results presented here hint that PSO has the competency to be integrated in choosing the aggregation sites in semi-decentralized federated learning. Nonetheless, further developments need to be done and studies to be conducted to ensure PSO's adaptation towards varying SDFL topologies and changing system characteristics.

\section{Related Work}
There are various propositions made to use PSO in task scheduling and load balancing, both in the cloud and the Edge. Below are a few most related works regarding placement optimization at the edge. 

One key aspect is computing offloading in Mobile Edge Computing (MEC). A study explored a Particle Swarm Optimization (PSO)-based task offloading strategy for 5G-enabled Industrial Internet of Things (IIoT) environments, optimizing energy efficiency and latency by distributing tasks among heterogeneous edge servers \cite{you2021efficient}. The PSO approach was compared with Genetic Algorithm (GA) and Simulated Annealing (SA), demonstrating its advantages in reducing latency and balancing energy consumption \cite{you2021efficient}.

Cloud computing task scheduling is another critical area. Researchers proposed a hybrid PSO-Genetic Algorithm (PSO-PGA) incorporating a phagocytosis mechanism to expand the search space and avoid local optima in cloud task scheduling \cite{fu2023task}. The phagocytosis mechanism, inspired by biological immune responses, allows weaker solutions to be engulfed and replaced by stronger ones, thereby maintaining diversity and preventing premature convergence. The study demonstrated improved completion times and convergence accuracy compared to traditional PSO and GA approaches \cite{fu2023task}.

Another study introduced a novel task scheduling approach in cloud computing using Dynamic Dispatch Queues (TSDQ) combined with hybrid meta-heuristic algorithms \cite{ben2018novel}. Two variations, one using Fuzzy Logic with PSO (FLPSO) and another integrating Simulated Annealing with PSO (SAPSO), were tested. The results indicated that FLPSO significantly reduced waiting time, queue length, makespan, and execution cost, beating other state-of-the-art scheduling strategies \cite{ben2018novel}.

Furthermore, edge aggregation and server placement in SDFL have been explored to address device association and resource allocation challenges. A study formulated an edge aggregation optimization problem and converted it into a dynamic optimization problem based on training loss degradation \cite{xu2024edge}. It introduced a Trilateral Matching-based Association (TMA) approach for efficient device association and resource allocation, which employs the classic Hungarian algorithm to derive the ideal matching set. Additionally, a Tabu Search-based Placement (TSP) approach was proposed to optimize the placement of edge servers. The combination of TMA and TSP in an iterative manner improved device participation reliability and edge aggregation efficiency \cite{xu2024edge}.

An adaptive PSO-based scheduling approach (AdPSO) was also proposed to optimize task execution in cloud computing \cite{nabi2022adpso}. This study introduced a new inertia weight strategy called Linearly Descending and Adaptive Inertia Weight (LDAIW) to improve the balance between local and global search. Experimental results showed that AdPSO achieved up to a 10 \% improvement in makespan, a 12 \% improvement in throughput, and a 60 \% improvement in resource utilization compared to existing PSO-based scheduling strategies \cite{nabi2022adpso}.


Overall, existing research provides various optimization techniques for task scheduling and offloading in edge and cloud environments. However, open challenges remain in balancing energy consumption, latency, and computational efficiency in SDFL systems, necessitating further exploration of hybrid meta-heuristic algorithms as black-box optimizers.


\section{Conclusion}
In this paper, we explored the usability of PSO as a black-box optimizer for aggregation placement in hierarchical semi-decentralized federated learning. We discussed that compared to other meta-heuristics, PSO shows faster and more accurate convergence. Our simulations and Docker-based implementations demonstrated that PSO efficiently optimizes client placement, reducing processing delay by balancing aggregation load across levels. We showed that PSO adapts well to large client numbers and outperforms random and uniform placement methods. Future work will explore adapting PSO for continuous system variations, adaptive particle sizes, and incorporating additional parameters into the fitness function. We will maintain PSO as a black-box solution and compare it with other meta-heuristic and learning-based approaches.



\end{document}